\documentclass[xcolor=pdftex,usenames,dvipsnames,table]{PoS}

\usepackage{multirow, lscape}
\usepackage{multicol}
\usepackage{array}
\usepackage{slashed}
\usepackage{graphicx}
\usepackage{subfig} 
\usepackage{amssymb,amsmath}
\usepackage{bm}
\usepackage{pifont}
\usepackage{tcolorbox}
\usepackage{colortbl}
\usepackage{chngpage}
\usepackage[export]{adjustbox}

\usepackage{url}
\title{Updates on Nucleon Form Factors from Clover-on-HISQ Lattice Formulation}

\ShortTitle{Nucleon Form Factors}

\author{\speaker{Yong-Chull Jang}\\
        Physics Department,
        Brookhaven National Laboratory, 
        Upton, NY, 11973, U.S.A.\\
        E-mail: \email{ypj@bnl.gov}}

\author{Tanmoy Bhattacharya, Rajan Gupta\\
        Theoretical Division, T-2, 
        Los Alamos National Laboratory, 
        Los Alamos, NM, 87545, U.S.A.}

\author{Huey-Wen Lin\\
        Department of Physics and Astronomy, 
        Michigan State University, MI, 48824, U.S.A}

\author{Boram Yoon\\
        Computer, Computational, and Statistical Sciences, CCS-7, 
        Los Alamos National Laboratory,
        Los Alamos, NM, 87545, U.S.A.}

\author{PNDME Collaboration}

\abstract{Updates on results for the electric, magnetic and axial
  vector form factors are presented. The data analyzed cover high
  statistics measurements on 11 ensembles generated with 2+1+1 flavors
  of HISQ fermions by the MILC collaboration. The data cover the range
  $0.057\fm < a < 0.15\fm$ in lattice spacing, $135\MeV < M_\pi < 320\MeV$ in the
  pion mass and $3.3 < M_\pi L < 5.5 $ in the lattice size. Fits to
  control excited-state contamination use up to 3-states in the
  spectral decomposition of the 3-point functions. The simultaneous
  chiral-continuum-finite-volume fits include leading order
  corrections in each of the three variables.  }

\FullConference{The 36th Annual International Symposium on Lattice Field Theory - LATTICE2018\\
		22-28 July, 2018\\
		Michigan State University, East Lansing, Michigan, USA.}

\graphicspath{{./figs/}}

\providecommand{\expv}[1]{\langle#1\rangle}

\providecommand{\MeV}{\;\mathrm{MeV}}

\providecommand{\fm}{\;\mathrm{fm}}
\renewcommand{\Re}{\;\mathrm{Re}}
\renewcommand{\Im}{\;\mathrm{Im}}

\begin{document}

\section{Introduction}
\label{sec:intro}

We present an update on the nucleon isovector electromagnetic and
axial vector form factors calculated using the HISQ-on-Clover lattice
formulation, which uses Wilson clover fermion action for valence
quarks and 2+1+1 flavors HISQ ensembles generated by the MILC
collaboration~\cite{Bazavov:2012xda}. The details of the lattice
calculation of form factors can be found in
Ref.~\cite{Bhattacharya:2013ehc,Rajan:2017lxk}, and details of the
lattice parameters for the 14 ensembles used to study form factors in
the published paper on charges~\cite{Gupta:2018qil}. To control
excited state contamination (ESC), we include up to 3-states in the
spectral decomposition of the 3-point
correlator~\cite{Jang:2018lup}.\looseness-1

Compared to the previous analysis of the axial vector form
factors~\cite{Rajan:2017lxk}, several updates have been made. First,
ensembles $a09m310$ and $a09m220$ now have higher statistics, an
addtional measurement with $\tau/a=16$ in the 3-point correlator, and
momentum insertion, $\bm{q}=2\pi\bm{n}/L$, with $\bm{n}^2 \leq
10$. Intermediate results with this update can be found in
Ref.~\cite{Jang:2018lup}. Second, all ensembles now use the truncated solver with
bias correction. Third, the two physical mass ensembles are updated:
higher statistics for $a06m135$ and new $a09m130W$ data with a wider
smearing at the source and sink to improve the overlap with the ground
state.  Fourth, $a06m310W$ and $a06m220W$ data with a wider smearing are
included. Fifth, added $a15m310$ at coarser lattice spacing $a \approx
0.15\fm$ and pion mass $M_\pi \approx 320$. Lastly, we investigate
finite volume effects by including three ensembles with different
volumes but the same lattice spacing $a=0.12\fm$ and pion mass
$M_\pi=220\MeV$.\looseness-1

\section{Electromagnetic Form Factors}
\label{sec:emff}

\begin{figure}[tbh!]
  \centering
  \includegraphics[valign=t,width=0.525\columnwidth]{%
    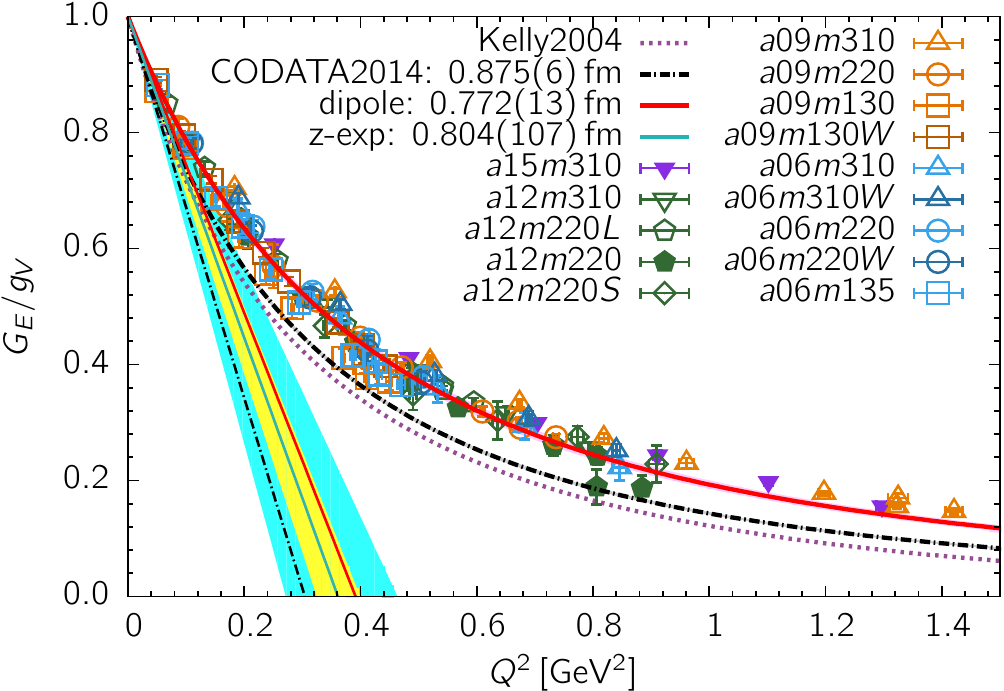}
    \hfill
  \includegraphics[valign=t,width=0.453\columnwidth]{%
    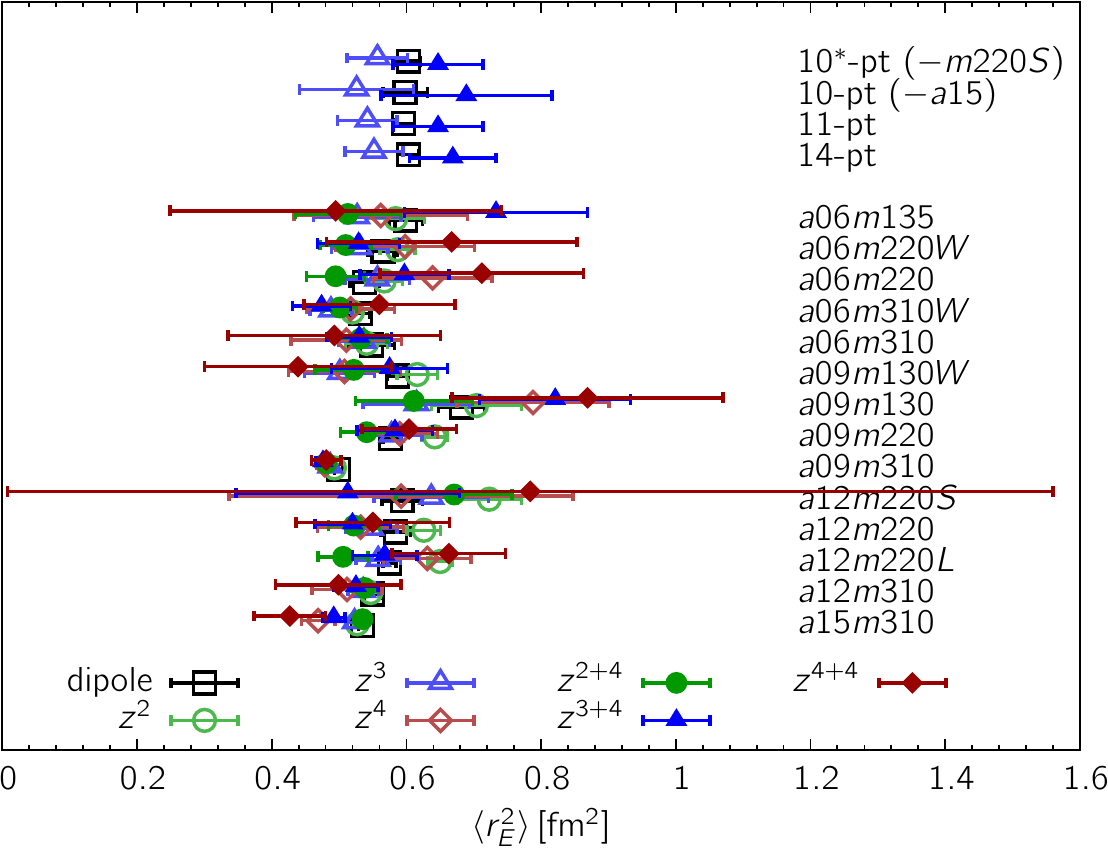}
  \caption{(left) Isovector electric form factor $G_E^{u-d}(Q^2)/g_V$
    compared to the Kelly parameterization, and dipole fit with CODATA
    value of $\expv{r_E}$.  The straight lines represent the slope at
    $Q^2=0$ for each fit. The yellow inner band and cyan outer band
    are the statistical and systematical errors in the
    $z$-expansion. (right) Values of $\expv{r_E^2}$ from the CCFV fits
    for the 14 calculations, each with the dipole and $z$-expansion
    analysis of the $Q^2$ behavior. }
  \label{fig:GE}
\end{figure}

\begin{figure}[tbh!]
\vspace{-4mm}
  \centering
  \includegraphics[valign=t,width=0.50\columnwidth]{%
    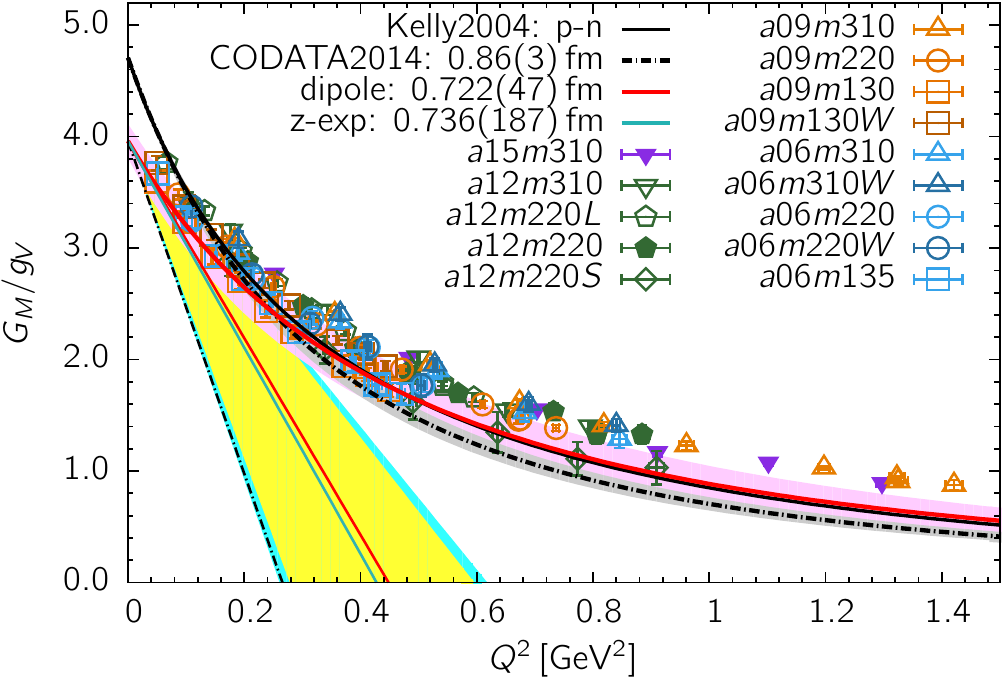}
    \hfill
  \includegraphics[valign=t,width=0.48\columnwidth]{%
    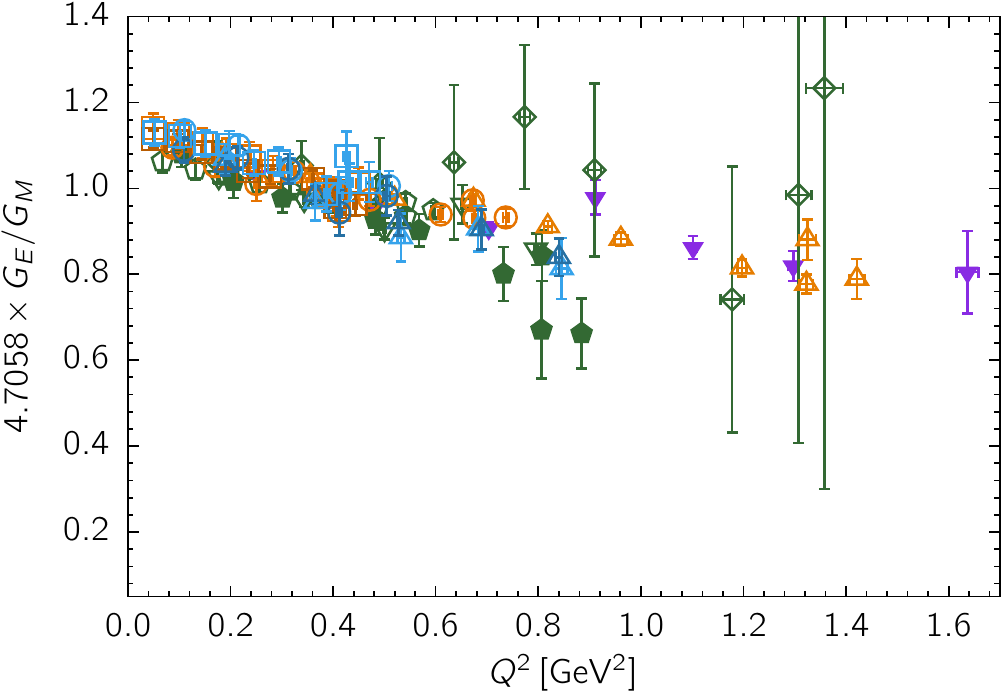}
    \caption{(left) Data for the isovector magnetic form factor
      $G_M^{u-d}(Q^2)$. To allow a visual comparison of the slope at
      $Q^2=0$, different values of $\mu=G_M(0)$ are shifted to a common
      point taken to be the dipole result. (right) Data for the ratio
      $\mu^\text{exp} G_E/G_M$ with $\mu^\text{exp}=4.7058$.}
    \label{fig:GM}
\end{figure}

Lorentz covariant decompsition of the matrix element of the
electromagnetic vector current $V_\mu$ between nucleon states can be written in
terms of Dirac, $F_1$, and Pauli, $F_2$, form factors as:
\begin{align}
\label{eq:VFFdef}
\left\langle N(\bm{p}_f) | V_\mu (\bm{q}) | N(\bm{p}_i)\right\rangle &=\; 
{\overline u}_N(\bm{p}_f)\left( F_1(Q^2) \gamma_\mu
+ \sigma_{\mu\nu} \frac{F_2(Q^2)}{2 M_N}\right) u_N(\bm{p}_i),
\end{align}
where $u_N(\bm{p})$ is the nucleon spinor, $M_N$ is the nucleon mass, and $Q^2=\bm{p}_f^2-(E-M_N)^2$ is the Euclidean four-momentum square transferred.
The analysis presented here is carried out in terms of the Sachs
electric and magnetic form factors $G_E$ and $G_M$:
%
\begin{align}
  G_E(Q^2) &= F_1(Q^2) - \frac{Q^2}{4M_N^2} F_2(Q^2) \,,\quad
  G_M(Q^2) = F_1(Q^2) + F_2(Q^2) \,.
  \label{eq:def-GEGM}
\end{align}
which are commonly used to express the Rosenbluth $ep$ cross-section.
Lattice data for $G_E$ and $G_M$ from all 14 calculations are summarized in
Figs.~\ref{fig:GE} and ~\ref{fig:GM}, and compared
with the Kelly parameterization for the isovector combination $p-n$ and 
with dipole fit using CODATA2014~\cite{Mohr:2015ccw} value for the proton
charge radius, $\sqrt{\expv{r_E^2}} = 0.875(6)$~fm. 

\begin{figure}[tb!]
  \vspace{-2mm}
  \centering
  \begin{minipage}[t]{0.26\textwidth}
    \includegraphics[width=0.92\columnwidth,valign=t]{%
    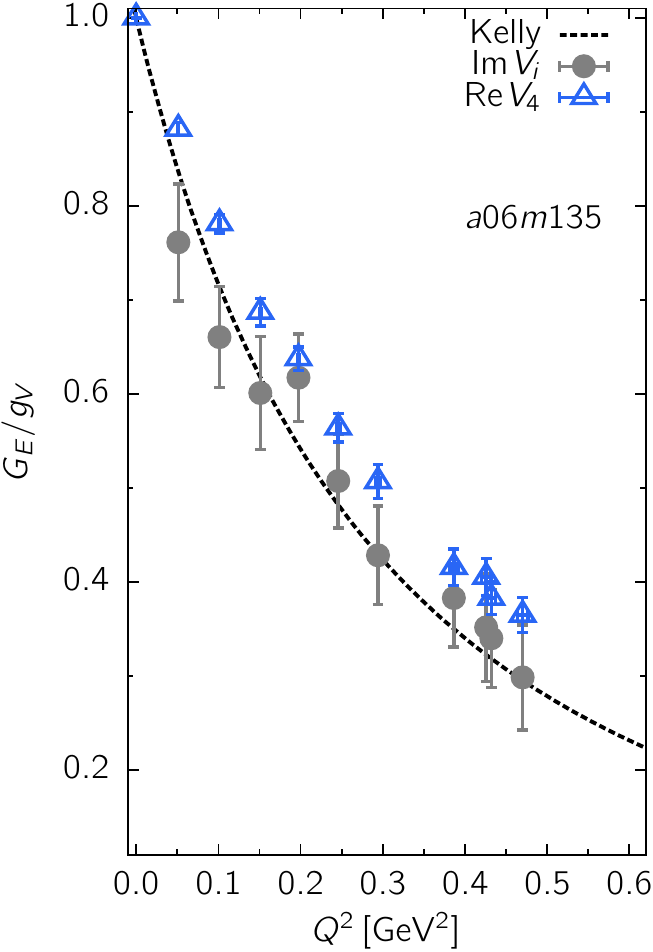}
  \end{minipage}
  \begin{minipage}[t]{0.72\textwidth}
    \includegraphics[trim=0 0 0 0,clip,width=0.48\columnwidth,valign=t]{%
      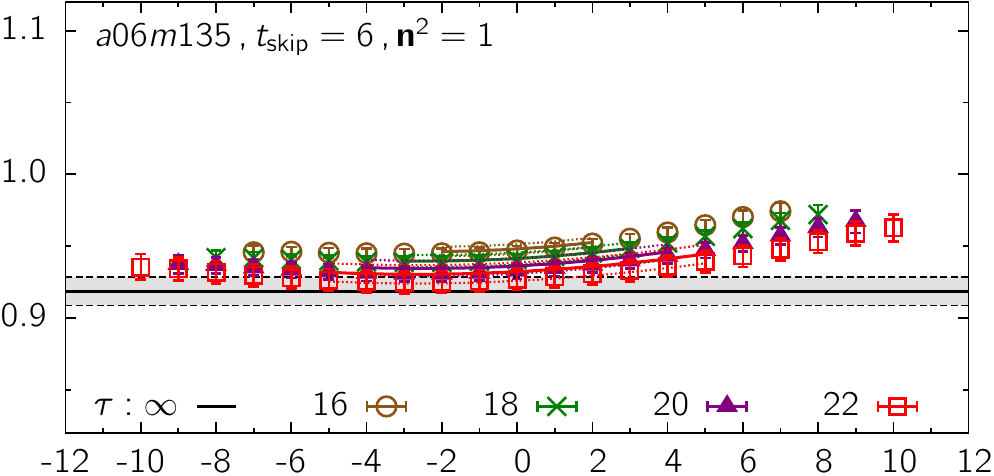}
    \includegraphics[trim=0 0 0 0,clip,width=0.48\columnwidth,valign=t]{%
      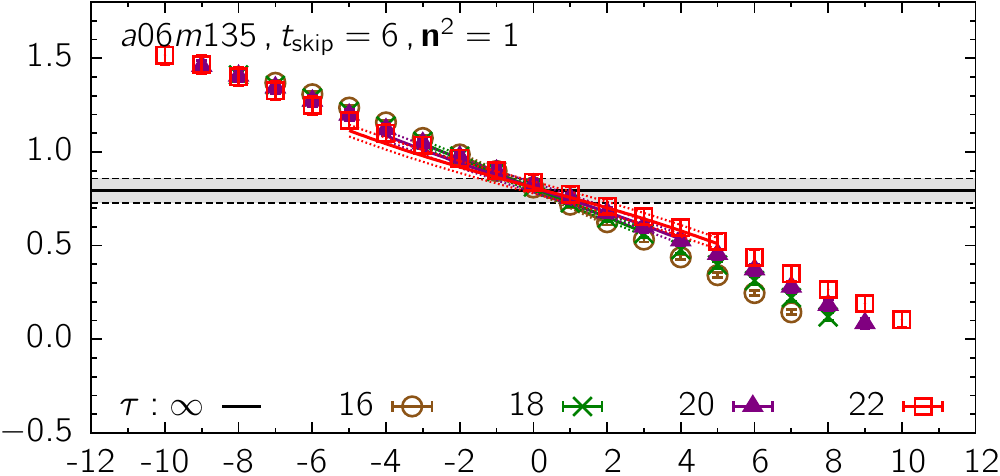}\\
      \hspace{0.25\textwidth}
    \includegraphics[trim=0 0 0 0,clip,width=0.48\columnwidth,valign=t]{%
      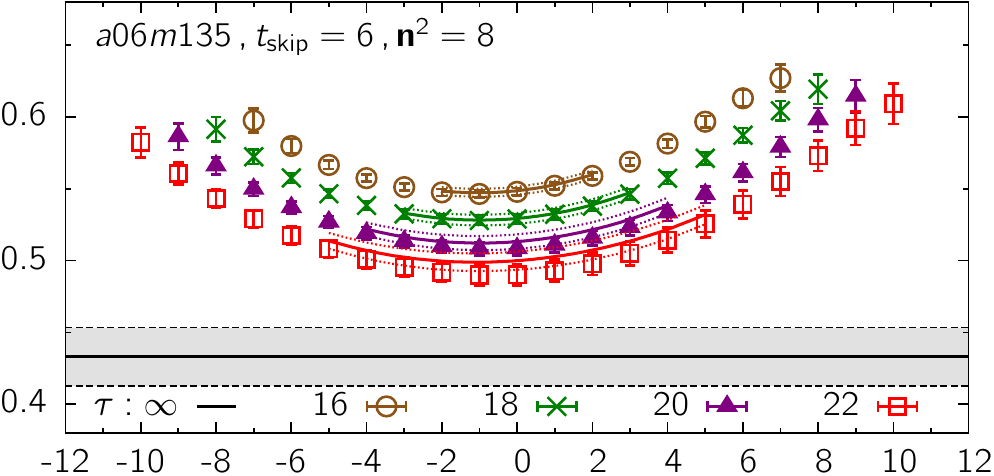}
    \includegraphics[trim=0 0 0 0,clip,width=0.48\columnwidth,valign=t]{%
      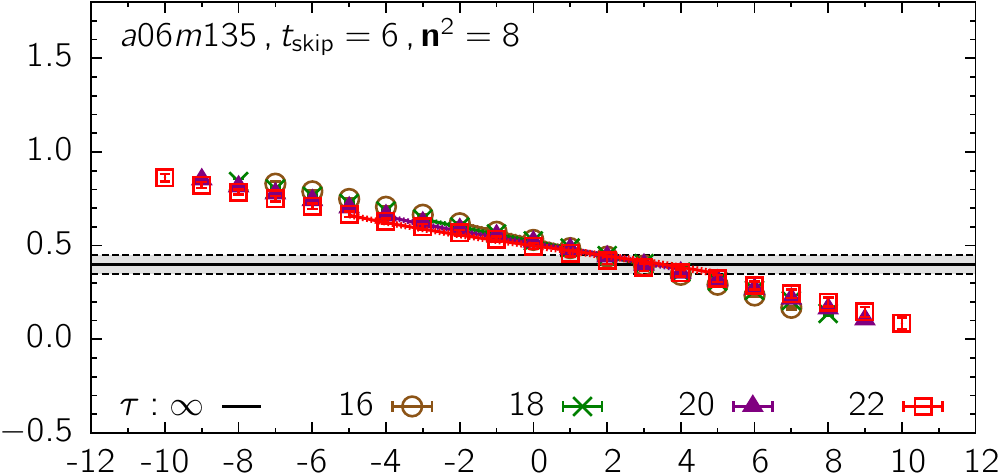}
  \end{minipage}
  \caption{(left) Comparison of $G_E(Q^2)$ from $\Im V_i$ versus $\Re
    V_4$. (middle) Excited state effects in matrix element of $\Re
    V_4$ and (right) $\Im V_i$. All data are from the physical mass
    ensemble $a06m135$.}
  \label{fig:GE-V4-Vi}
\end{figure}

Our final results for $ G_E(Q^2)$ are taken from the $\Re V_4$
component of the current versus $\Im V_i$ averaged over $i=1,2,3$.  A
comparison of the two estimates is shown in Fig.~\ref{fig:GE-V4-Vi}
for the $a06m135$ ensemble along with examples of the pattern of ESC. 
The lack of a plateau in the matrix element of $\Im V_i$ gives rise to
a much larger error than in $\Re V_4$.
The error and difference increases as $Q^2$, $a$ and $M_\pi$
decrease. The reasons for these (discretization errors, ESC, finite
volume effects) are not understood. For present, we take $\expv{r_E^2}$
from $\Re V_4$ since precise values of $G_{E}$ at small $Q^2$ are needed
to determine the electric charge radius defined as $\expv{r_E^2} =
-6\frac{d}{dQ^2} \frac{G_E(Q^2)}{g_{V}}\big\rvert_{Q^2=0}$. 
%
%

The continuum-chiral-finite-volume (CCFV) fits for the electric and
magnetic charge radii, $\expv{r_E^2}$ and $\expv{r_M^2}$, and magnetic
moment, $\mu$, are carried out including the leading order terms that
describe lattice artifacts due to finite lattice spacing $a$, and
variation with pion mass $M_\pi$ and finite volume parameter $M_\pi L$
using
expressions taken from
Refs.~\cite{Beane:2004tw,Gockeler:2003ay,Bernard:1998gv}.

\begin{equation}
  \expv{r_E^2}(a,M_\pi,L) = c_1^E + c_2^E a + c_3^E \ln(M_\pi^2/\lambda^2) + c_4^E \ln(M_\pi^2/\lambda^2) \exp(-M_\pi L) \,,
  \label{eq:rEsq-extrap}
\end{equation}
\begin{equation}
  \expv{r_M^2}(a,M_\pi,L) = c_1^M + c_2^M a + c_3^M/M_\pi + c_4^M/M_\pi \exp(-M_\pi L)\,, 
  \label{eq:rMsq-extrap}
\end{equation}
\begin{equation}
  \mu(a,M_\pi,L) = c_1^\mu + c_2^\mu a + c_3^\mu M_\pi + c_4^\mu M_\pi\left(1-\frac{2}{M_\pi L}\right) \exp(-M_\pi L) \,. 
  \label{eq:mu-extrap}
\end{equation}

\begin{figure}[tbh!]
\vspace{-2mm}
  \centering
    \includegraphics[trim=20 0 0 0,clip,width=0.32\columnwidth]{%
    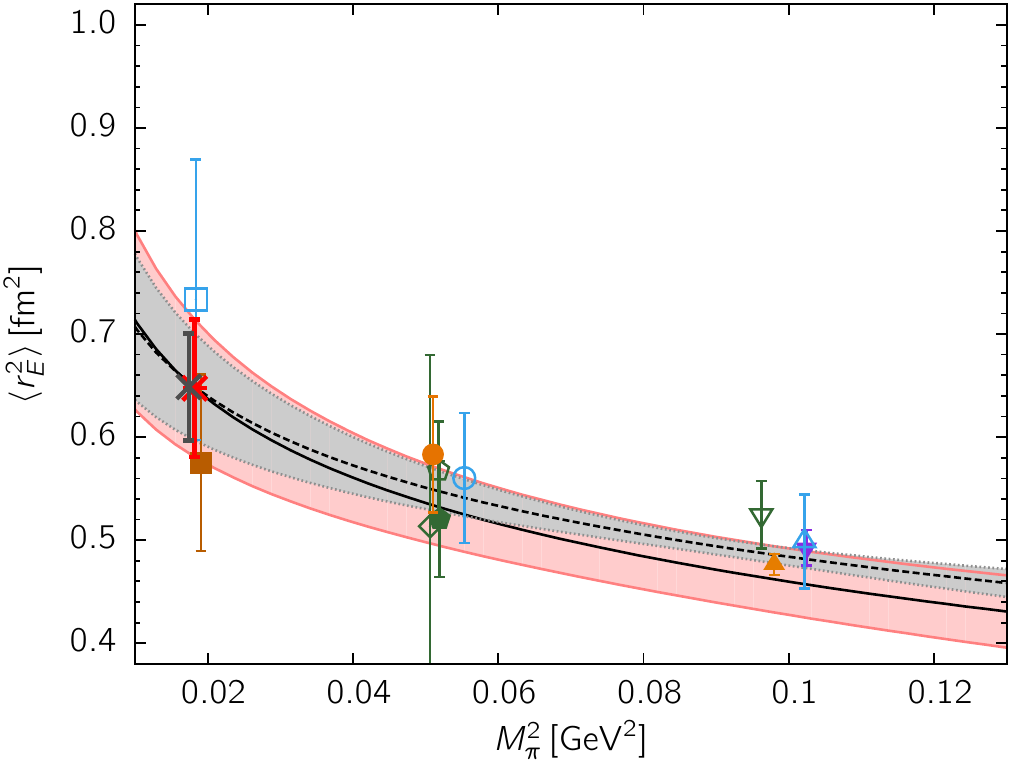}
    \includegraphics[trim=20 0 0 0,clip,width=0.32\columnwidth]{%
    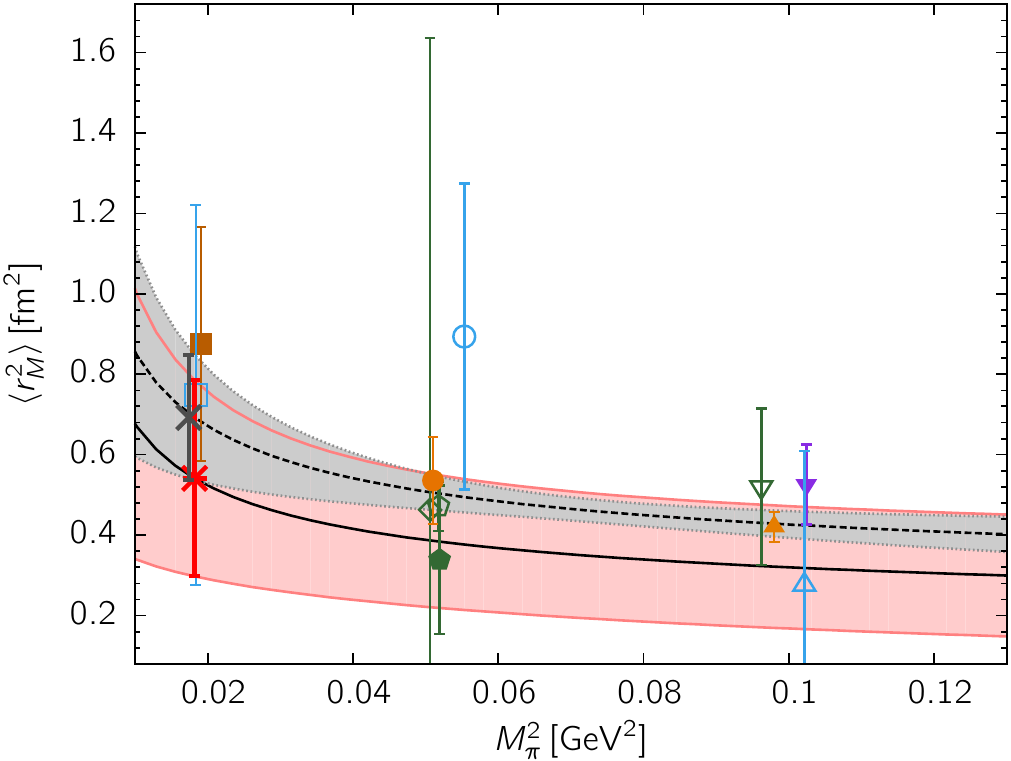}
    \includegraphics[trim=18 0 0 0,clip,width=0.32\columnwidth]{%
    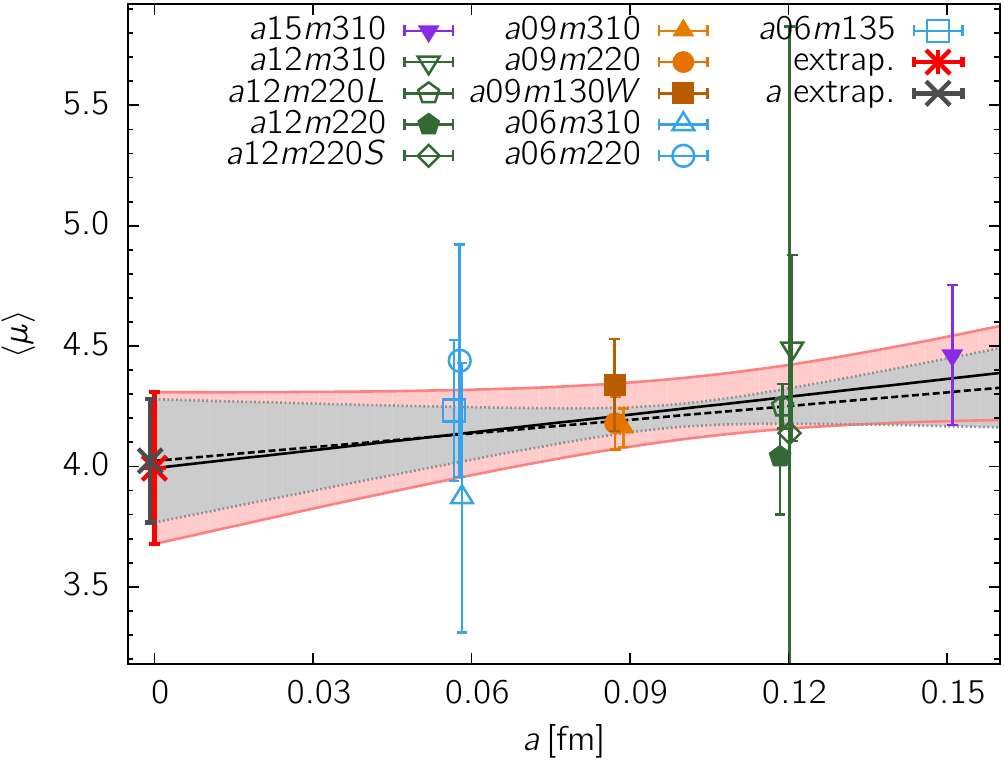}
    \caption{CCFV fits (pink band) plotted versus the variable they
      are most sensitive to. Fits to $\expv{r_E^2}$ (left) and
      $\expv{r_M^2}$ (middle), both in $\fm^2$, highlight the
      non-analytical dependence on $M_\pi$ given in
      Eqs.~\protect\eqref{eq:rEsq-extrap}
      and~\protect\eqref{eq:rMsq-extrap}. (right) CCFV fit for
      magnetic moment $\expv{\mu}$ is plotted versus $a$. In contrast,
      the gray band in each panel is the result of fits versus the
      single variable defining the x-axis.}
    \label{fig:ccfv-em}
\end{figure}

We first analyze the $Q^2$ dependence of form factors for each of the
14 calculations using the dipole and model independent
$z$-expansions. Then recognizing that the two pairs, $a06m220(W)$ and
$a06m310(W)$, share the same gauge ensemble but only the smearing
widths are different, we construct 11-point data by averaging the
$\expv{r_M^2}$, $\expv{r_M^2}$ and $\mu$ values from the two 
smearings assuming full correlations, and dropping the $a09m130$ data
because the bias correction is not available for $Q^2\neq 0$. We further
construct two 10-point data sets by excluding (i) the coarsest
lattice spacing $a15m310$ or (ii) the smallest volume $a12m220S$ data. The
CCVF fits for getting the final $\expv{r_E^2}$, $\expv{r_M^2}$, and
$\mu$, are then performed for for each of the data sets: 14-point,
11-point, 10-point and $10^\ast$-point and for the dipole and various
$z$-expansion analysis of $Q^2$ behavior.  In these fits, the 14, 11,
10 or $10^\ast$ data points are treated as uncorrelated.

The variation of $\expv{r_E^2}$ between the 14 calculations and
between dipole and $z$-expansion analysis is small as shown in
Fig.~\ref{fig:GE}. The figure also shows the results from the 14-,
11-, 10-, and 10$^\ast$-point CCFV fits. We take the central value for
$\expv{r_E^2}$ from the 11-point fit with the $z^{3+4}$ truncation of
the $z$-expansion. This is given in Table~\ref{tab:summary}. Note that
$z^{i+4}$ fits include the four sum rule constraints imposed to ensure
that $G_{A,E,M}(Q^2) \to 0$ as $1/Q^4$ for large $Q^2$. For
$\expv{r_M^2}$ and $\mu$, we take the central values from $z^3$ as the
$z^{3+4}$ fits are unstable. The reason $G_M$ fits are less stable is
the point $G_M(0)$, which would pin the fit at $Q^2=0$ is not
calculable from lattice simulations.
The CCFV fits for $\expv{r_E^2}$, $\expv{r_M^2}$, and
$\expv{\mu}$ are shown in Fig.~\ref{fig:ccfv-em} versus the variable they 
vary the most with. \looseness-1

The magnetic moment $\mu$ from either the dipole or the $z$-expansion
analysis is about $15\%$ smaller than the experimental value
$\mu^\text{exp}=4.7058$ as can be inferred from the data for
$\mu^\text{exp} G_E/G_M$ plotted in Fig.~\ref{fig:GM}.  Since the
ratio $G_E/G_M$ is insensitive to $a$ or $M_\pi$, the low value of
$\mu$ obtained is not simply explained by the less stable fits to
$G_M$ versus $Q^2$.  Also, the size and quality of the ESC in $\Re
V_i$ (averaged over $i=1,2$ as these two are related by the lattice
rotational symmetry) is similar to $\Re V_4$ shown in
Fig.~\ref{fig:GE-V4-Vi}, and the 2- and $3^\ast$-state fits give
stable estimates of the $\tau \to \infty$ value.

\section{Axial Form Factors}

\begin{figure}[tbh!]
  \centering
    \includegraphics[trim=0 0 0 0,clip,width=0.355\columnwidth]{%
    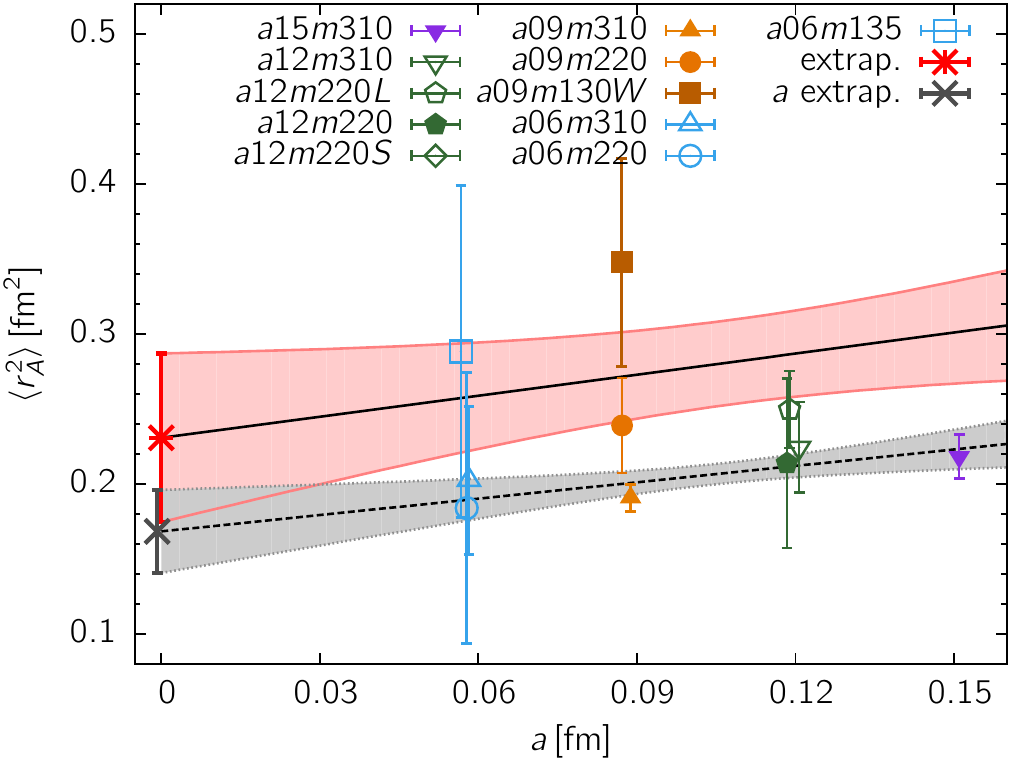}
    \includegraphics[trim=38 0 0 0,clip,width=0.31\columnwidth]{%
    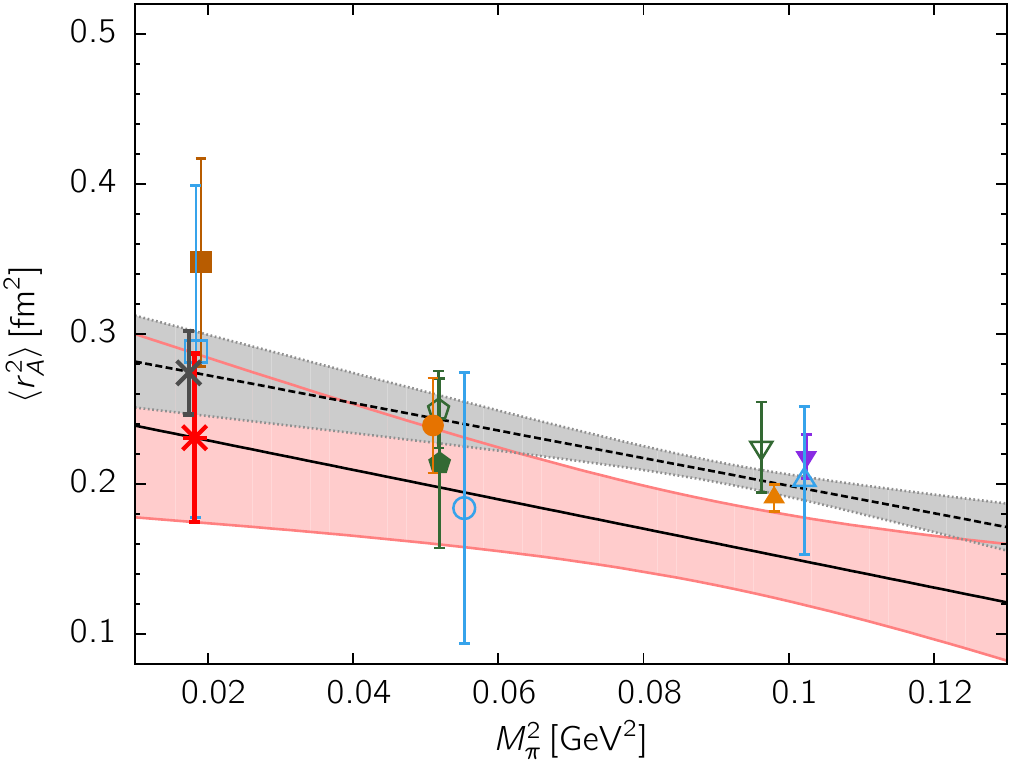}
    \includegraphics[trim=38 0 0 0,clip,width=0.31\columnwidth]{%
    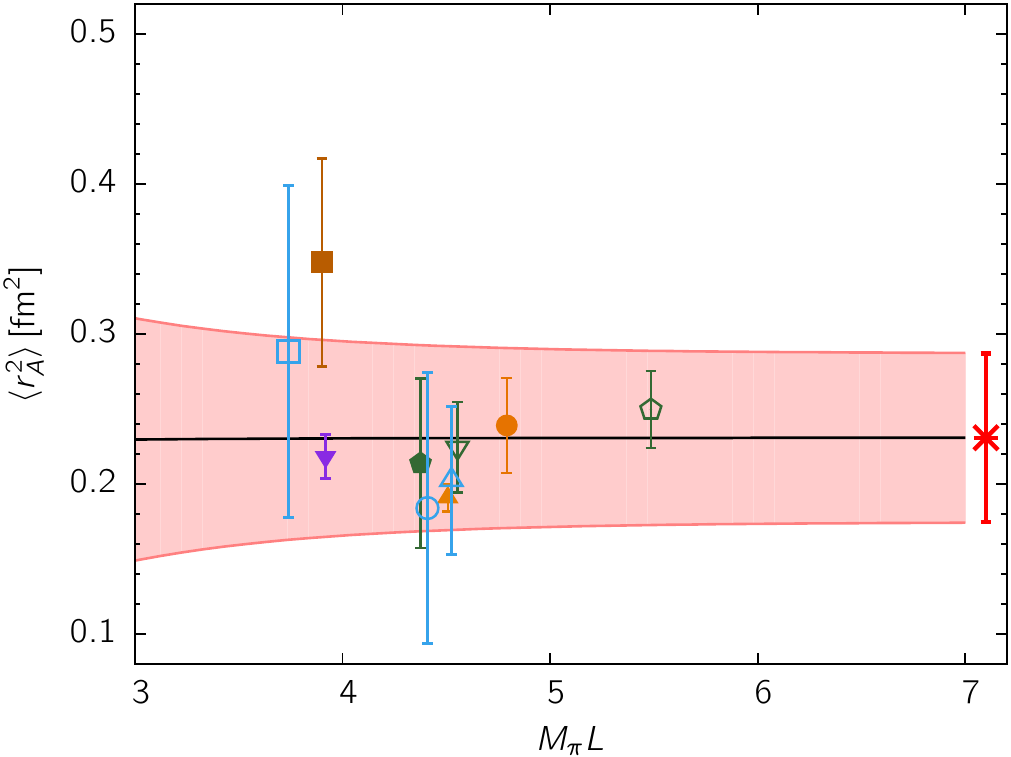}
    \caption{CCFV fits for the axial charge radius $\expv{r_A^2}$ in
      fm${}^2$. The three panels show the fits versus the lattice
      spacing (left) pion mass (middle) and finite box size 
      parameter $M_\pi L$ (right), with the other variables in 
      Eq.~\protect\eqref{eq:rAsq-extrap} set to their 
      physical values.}
    \label{fig:ccfv-aff}
\end{figure}

\begin{figure}[tbh!]
\vspace{-2mm}
  \centering
  \includegraphics[valign=t,width=0.51\columnwidth]{%
    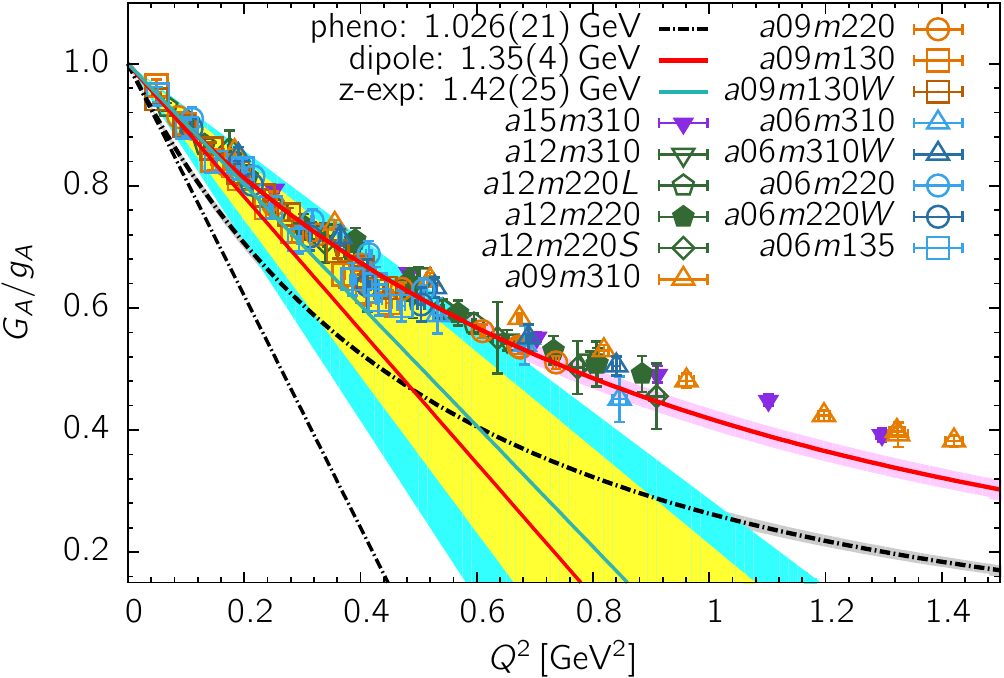}
    \hfill
  \includegraphics[valign=t,width=0.43\columnwidth]{%
    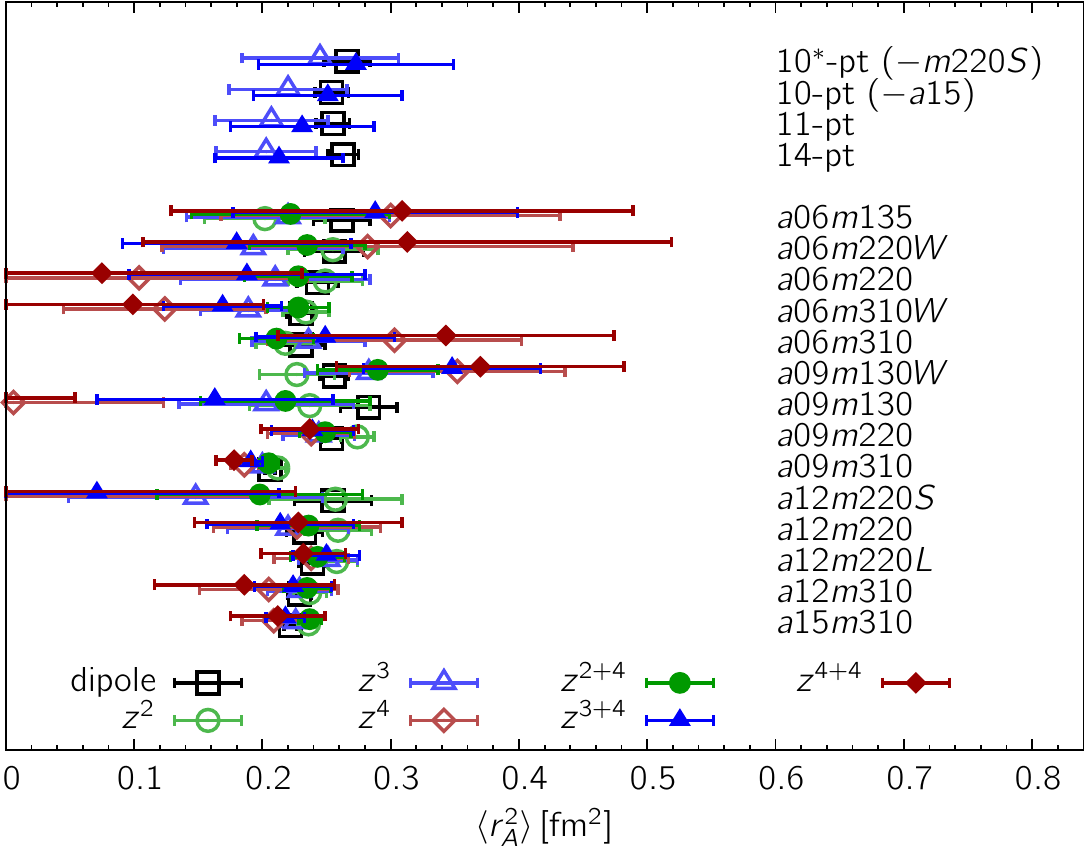}
  \caption{(left) Data for the axial-vector form factor
    $G_A^{u-d}(Q^2)/g_A$. The straight lines show the slope at $Q^2=0$
    for the dipole and $z$-expansion fits. The dashed-dotted line is
    the dipole fit with the phenomenological estimate
    $M_A=1.026$. (right) The values of $\expv{r_A^2}$ from the
    four CCFV fits, and from the 14 calculations with both the dipole
    and $z$-expansion analysis of the $Q^2$ behavior.}
  \label{fig:GA}
\end{figure}

The form factor decomposition of the matrix element of the 
isovector axial-vector current between nucleon states is: 
\begin{align}
\label{eq:AFFdef}
\left\langle N(\bm{p}_f) | A_\mu (\bm{q}) | N(\bm{p}_i)\right\rangle &=\; 
{\overline u}_N(\bm{p}_f)\left( G_A(Q^2) \gamma_\mu
+ q_\mu \frac{\tilde{G}_P(Q^2)}{2 M_N}\right) \gamma_5 u_N(\bm{p}_i) \,.
\end{align}
We follow the same procedure for controlling the ESC in the matrix
elements and for the extraction and CCFV fits to the axial charge radius as for
the electromagnetic form factors described in 
Sec.~\ref{sec:emff}.  The CCFV fit for $\expv{r_A^2}$
is made using 
\begin{equation}
  \expv{r_A^2}(a,M_\pi,L) = c_1^A + c_2^A a + c_3^A M_\pi^2 + c_4^A M_\pi^2 \exp(-M_\pi L) \,,
  \label{eq:rAsq-extrap}
\end{equation}
and results for the $z^{3+4}$ analysis are shown in
Fig.~\ref{fig:ccfv-aff}. All the data versus $Q^2$ are shown in
Fig.~\ref{fig:GA} (left), and the variation versus different fit
ansatz is shown in Fig.~\ref{fig:GA} (right). The central values are
taken from the 11-point CCFV fit and results for the dipole and
$z^{3+4}$ analysis are given in Tab.~\ref{tab:summary}.
%

\section{Discussion and Outlook}

Results for the charge radii $\sqrt{\expv{r_E^2}}$,
$\sqrt{\expv{r_M^2}}$, $\sqrt{\expv{r_A^2}}$ and magnetic moment $\mu$
from isovector electromagnetic and axial form factors are summarized
in Tab.~\ref{tab:summary}. The $z$-expansion results have larger
errors in all cases, and the dipole results are consistant with these
within statistical errors.
Compared to the previous works~\cite{Rajan:2017lxk,Jang:2018lup}, the
errors in the dipole estimates are smaller with increased statistics
and ensembles, but only in $r_E$ for the $z$-expansion.  Part of the
reason is that in Refs.~\cite{Rajan:2017lxk,Jang:2018lup}, the
$z$-expansion results were averaged: $z^2$ and $z^3$ for
$\expv{r_M^2}$ and $\mu$, and $z^{2+4}$ and $z^{3+4}$ for
$\expv{r_E^2}$ and $\expv{r_A^2}$. As a result, these errors quoted
were dominated by the smaller error points $z^2$ or $z^{2+4}$, while
the new results are taken from the higher order truncation, $z^3$ or
$z^{3+4}$, that have larger errors.  The more pressing challenge with
the $z$-expansion analysis is to show stability with respect to the
order of the truncation.

We now also quote a systematic error from the CCFV fits. For
${\expv{r_E^2}}$ and ${\expv{r_M^2}}$, the dominant variation is with
respect to $M_\pi^2$, so we take it to be the difference between the
two physical pion mass ensemble results. For the magnetic moment, the 
largest variation is with $a$, so we take the difference between the 
CCFV fit result and the average of the five finest lattice, $a\approx 0.06\fm$, 
results. These conservative estimates will be refined in future work.

Adding the statistical and systematic errors in quadrature, our
lattice estimates for $r_E$ are consistent with the Kelly
parameterization of the experimental data for the isovector
combination. Clearly, the current precision in the lattice data is not
sufficient to address the proton charge radius puzzle. The lattice
estimate of the magnetic charge radius $r_M$ has an even larger
error. The magnetic moment $\mu$ from our calculation undershoots the
experimental value by $15\%$, with the $a \approx 0.06\fm$ data
pulling down the CCFV fit. The range of parameter values analyzed in
this work leaves open the possibility that the finite volume effects
are significant, especially at the lowest $Q^2$
value~\cite{Tiburzi:2007ep}.  For the future, better control over
extrapolation to the physical limit would include a modified CCFV fit
that includes higher order corrections from an effective theory such
as HB$\chi$PT~\cite{Bernard:1998gv}, and by combining this fit with
the $Q^2$ behavior as discussed in Ref.~\cite{Bali:2018qus}.

The axial charge radius $r_A$ is smaller than the value extracted from
from neutrino scattering data, $r_A=0.666(17)\fm$, from
electroproduction, $r_A=0.639(10)\fm$, and from a reanalysis of
deuterium data, $r_A=0.68(16)\fm$. In Ref.~\cite{Rajan:2017lxk}, we
had pointed out a problem with the lattice estimates of the axial
$G_A$, induced pseudoscalar $\widetilde{G}_P$, and pseudoscalar $G_P$
form factors: while the PCAC relation is satisfied at the 
correlator level, it is not satisfied by the form factors. This 
problem is under investigation.

\begin{table}[tbh!]
  \centering
  \renewcommand{\arraystretch}{0.9}
  \resizebox{.95\textwidth}{!}{%
  \begin{tabular}{l|l|l|ll|ll}
  \hline\hline
    & & $r_E\,[\mathrm{fm}]$ & $r_M\,[\mathrm{fm}]$ & $\mu$ & $r_A\,[\mathrm{fm}]$ & $\mathcal{M}_A\,[\mathrm{GeV}]$ \\
    \hline
    \multirow{2}{*}{This work} & $z$-exp. & 0.804(42)(98) & 0.736(166)(86) & 3.99(32)(17) & 0.481(58)(62) & 1.42(17)(18) \\
    & dipole   & 0.772(10)(8)  & 0.722(23)(41)  & 3.96(10)(12) & 0.505(13)(6) & 1.35(3)(2) \\
    \hline
    \multirow{2}{*}{Jang~{\it~et.~al.}\cite{Jang:2018lup}} & $z$-exp. & 0.83(9) & 0.82(10) & 3.47(36) & 0.50(6) &  1.36(17) \\
    & dipole   & 0.79(3)  & 0.77(4)  & 3.72(23) & 0.51(2) & 1.34(6)  \\
    \hline
    \multirow{2}{*}{Gupta~{\it~et.al.}\cite{Rajan:2017lxk}} & $z$-exp. & & & & 0.46(6) &  1.48(19) \\
    & dipole   & & & & 0.49(3) & 1.39(9)  \\
  \hline\hline
  \end{tabular}}
  \caption{Summary of charge radii, magnetic moment, and axial mass
    from isovector form factors. When two errors are given, the first is
    statistical and the second is systematic.}
  \label{tab:summary}
\end{table}

\section*{Acknowledgments}
We thank the MILC collaboration for sharing the 2 + 1 + 1-flavor HISQ ensembles generated by them. We gratefully acknowledge the computing facilities at and resources provided by NERSC, Oak Ridge OLCF, USQCD and LANL Institutional Computing.

%
\bibliography{lattice2018}

\end{document}